%% file: texfiles/main.tex
\documentclass[runningheads]{llncs}
\usepackage[T1]{fontenc}
\usepackage{graphicx}
\usepackage{makecell}
\usepackage{graphicx}
\usepackage{array}
\usepackage{booktabs}
\usepackage[sort&compress]{natbib}

\bibpunct{[}{]}{,}{n}{,}{,}
\newsavebox\mysavebox
\begin{document}

\title{Reimagining the Augmented Reality Accessibility Ecosystem for Deaf Students: Service Provider Perspectives in Experiential Learning}

\titlerunning{Service Provider Perspectives on AR Accessibility for Deaf Students}

\author{Roshan Mathew\inst{1}\thanks{Corresponding author: rm1299@rit.edu}\orcidID{0000-0001-6618-2802} \and
Roshan Peiris\inst{1}\orcidID{0000-0002-4191-3565}}

\authorrunning{R. Mathew and R. Peiris}

\institute{Accessible and Immersive Realities Lab \\ Rochester Institute of Technology, Rochester, NY, USA \\
\email{rm1299@rit.edu, roshan.peiris@rit.edu}
}

\maketitle              
\begin{abstract}
In experiential learning environments, Deaf and Hard of Hearing (DHH) students often experience \textit{“split attention,”} dividing their focus among tasks, instructors, and access providers. Augmented Reality (AR) has been proposed as a means to centralize communication access within the student’s field of view; however, little is known about how such systems affect the instructors, interpreters, and captioners who support access in these settings. We present a formative, expert-based evaluation of ARRAE, an AR-mediated communication access ecosystem, examining the experiences of an instructor, an American Sign Language (ASL) interpreter, and a real-time captioner in a simulated laboratory environment. Comparing in-person, traditional remote, and AR-mediated access, our findings suggest that AR reconfigures accessibility labor and interactional practices by redistributing communication, attention, and awareness across participants. While AR-mediated access supported integrated, in-situ communication and enhanced contextual awareness, it also altered interactional feedback mechanisms, such as gaze coordination and pacing, that support shared understanding during instruction. These observations highlight trade-offs between access, control, and coordination, and suggest that AR-mediated systems may be most effective when considered as part of a broader, hybrid ecosystem rather than as standalone replacements for existing modalities. This work contributes initial insights into how AR reshapes communication access in experiential learning and informs the design of systems that support coordinated, multi-user interaction in hands-on educational contexts. 

\end{abstract}

\input{texfiles/1-introduction}

\input{texfiles/2-related-works}
\input{texfiles/3-method}
\input{texfiles/4-findings}
\input{texfiles/5-discussion}

\input{texfiles/6-limitations-futurework}
\input{texfiles/7-conclusion}
\input{texfiles/8-credits}

\bibliographystyle{splncs04}
\bibliography{bib/main.bib}

\end{document}

%% file: texfiles/1-introduction.tex
\section{Introduction}
Accessibility in higher education is typically framed as a service delivered to a student. However, in practice, it is a complex orchestration of labor involving multiple stakeholders: the instructor managing the pedagogy, the interpreter or captioner translating the content, and the student synthesizing it. This coordination becomes fragile in experiential learning environments, such as chemistry laboratories, where visual attention is a scarce resource. In these settings, DHH students face the \textit{"split-attention effect,"} forcing a choice between looking at their experimental task, a potential safety hazard, and an interpreter \cite{smithchemical2016}. While recent work has explored Augmented Reality (AR) smart glasses to overlay access services directly into the student’s field of view \cite{mathew_access_2022, miller_use_2017}, these interventions are almost exclusively evaluated through the lens of the end user's user experience. However, access technologies do not exist in a vacuum. If a system reduces the student's cognitive load but increases the cognitive burden on the interpreter or disconnects the instructor from their instructional rhythm, the ecosystem fails.

Our work presents a formative, expert-based evaluation of ARRAE (Augmented Reality Real-Time Access for Education) that centers the accessibility provider ecosystem. Through an analysis of the experiences of an instructor, an American Sign Language (ASL) interpreter, and a real-time captioner in a simulated laboratory setting, we make the following contributions: (1) A qualitative account of how AR-mediated communication access reshapes accessibility labor and interactional practices among instructors, interpreters, and captioners during hands-on laboratory tasks, in comparison to in-person and conventional remote access; and (2) An articulation of key usability, workflow, and design considerations that emerge when deploying AR-mediated communication access in experiential higher education settings.

%% file: texfiles/2-related-works.tex

%% file: texfiles/3-method.tex
\section{Method}

\subsection{ARRAE Platform Implementation}
We developed ARRAE, a platform that connects support personnel to deaf students wearing optical see-through (OST) smart glasses \cite{vuzix_blade2}. The platform provides distinct interfaces for students and support personnel. Students use a streamlined heads-up display (HUD) to receive live video of the interpreter or captioning within their line of sight, while support personnel use a web-based interface to manage real-time access services.

\subsubsection{Access for DHH Students}
For deaf students, ARRAE functions as a receiver, overlaying either a live video feed of an interpreter or a scrolling text stream of captions directly into the student's line of sight. This design follows the spatial contiguity principle \cite{mayer_spatial_contiguity_principle_2009}, allowing students to remain visually engaged with the task while receiving access to spoken instruction.

\subsubsection{The Support Ecosystem}
Unlike traditional remote access solutions, where support personnel may see only a limited webcam view of the student, ARRAE instruments the lab with three Pan-Tilt-Zoom (PTZ) cameras: a Workstation Camera (workbench view), an Observation Camera (wide-angle view of the student's signing space), and an Instructor Camera (instructor view). These feeds are streamed via WebRTC to a browser-based dashboard with sub-second latency.

\paragraph{Interpreting Workflow.}
Interpreters broadcast their video feed directly to the student's smart glasses display. Their dashboard shows all active lab feeds simultaneously, allowing interpreters to monitor both the student's workbench and signing space to support context-aware interpretation. The architecture also supports team interpreting standards \cite{rid_resources_2022, hoza_team_2022}.

\paragraph{Captioning Workflow.}
Captioners operate through a view integrated with a TypeWell ingestion server \cite{typewell_transcription}. This interface supports rapid text entry and delivery while also displaying the PTZ camera feeds, providing visual context for task-relevant actions that may not be evident from audio alone.

\paragraph{Instructor Monitoring.}
While students view only their selected access service to reduce cognitive load, the instructor's dashboard aggregates all active feeds, supporting situational awareness during hands-on experiments.

\subsection{Study Design}
We conducted a formative, expert-based evaluation of ARRAE in a simulated laboratory setting. The study was approved by the institutional review board, and all participants provided informed consent. The broader study involved 12 deaf student participants: six in the interpreting study and six in the captioning study. Deaf student perspectives are reported in a separate study \cite{mathew_experiential_2026}. This paper focuses on the qualitative experiences of three expert participants: one STEM instructor (10--15 years of experience), one certified ASL interpreter (3--5 years of VRI experience), and one professional captioner (5--10 years of real-time captioning experience). The instructor participated in both studies, which were run separately; the interpreter participated only in the interpreting study, and the captioner only in the captioning study.

An expert review approach was used to identify interactional barriers, workflow demands, and role-specific design requirements relevant to early-stage system development. Three service delivery conditions were compared: (1) In-Person Access, (2) Remote Access (traditional VRI/captioning via tablet), and (3) AR-Mediated Access (ARRAE).

\subsubsection{Study Environment and Instrumentation} 
To simulate remote service delivery, the study utilized two distinct locations. 

\begin{figure*} \centering \setlength{\belowcaptionskip}{-15pt} \includegraphics[width=0.8\textwidth]{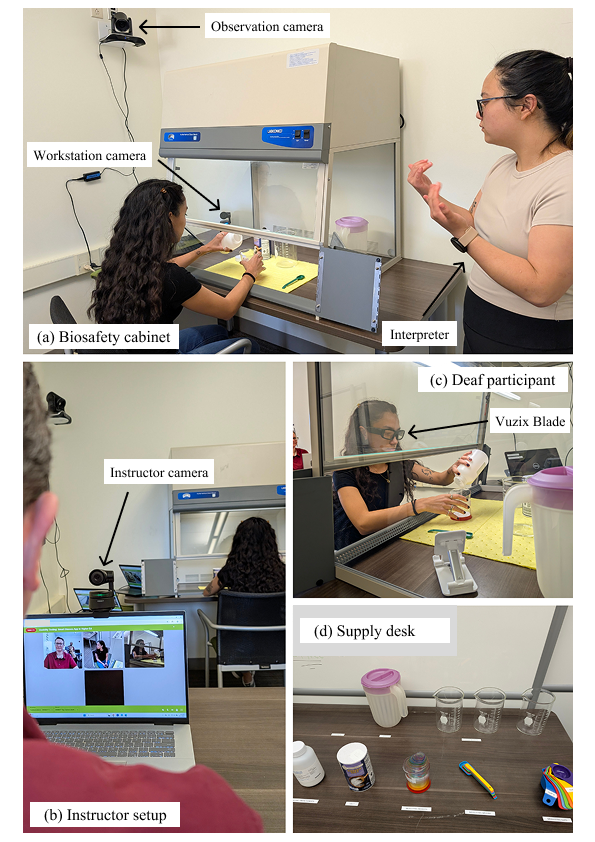} \caption{ARRAE study setup at Location A. (a) Biosafety cabinet with workstation and observation cameras. (b) Instructor setup. (c) Deaf participant wearing Vuzix Blade smart glasses. (d) Supply desk for student participants.} \label{fig:location-a} \end{figure*}

\paragraph{Location A: Simulated Lab.}
A workstation with a biosafety cabinet was used to approximate a STEM laboratory environment (Fig.~\ref{fig:location-a}). The space was instrumented with an OBSBOT Tail Air 4K PTZ camera \cite{obsbot_tail_air} for the workbench, an overhead Tenveo 10x Zoom PTZ camera \cite{tenveo_vhd10u} for the student's wider signing space, and an OBSBOT Tiny 4K webcam \cite{obsbot_tiny} for the instructor. All feeds were streamed to the ARRAE dashboard. PTZ cameras were chosen based on the findings from a prior study where interpreters reported a preference for PTZ cameras in remote interpreting contexts \cite{mathew_improving_2025}. 

\begin{figure*} \centering \setlength{\belowcaptionskip}{-10pt} \includegraphics[width=0.8\textwidth]{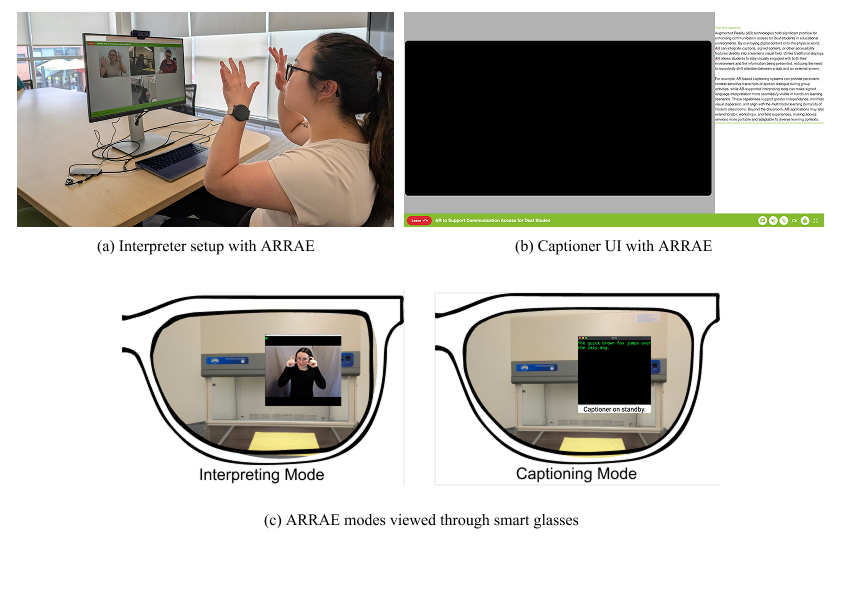} \caption{ARRAE interfaces. (a) Interpreter setup at Location B. (b) Captioner user interface. (c) Smart glasses view illustrating interpreting and captioning modes.} \label{fig:location-b} \end{figure*}

\paragraph{Location B: Remote Support Hub.}
During the Remote and ARRAE conditions, the interpreter and captioner worked from a separate room equipped with a laptop and external monitor (Fig.~\ref{fig:location-b}), allowing them to view the lab feeds and stream interpretation or captions back to the student.

\subsubsection{Procedure}
The interpreting and captioning studies were conducted separately. Each study included six sessions with six different deaf student participants, segregated according to their preference for the primary access modality. In each session, the student completed three hands-on laboratory tasks (making artificial snow using an instant snow polymer) guided by spoken English instructions, experiencing all three access conditions. Condition order and task order were counterbalanced across sessions. Sessions were moderated by deaf researchers, and students received smart-glasses training before beginning the tasks. The same instructor participated across all conditions in both studies.

\subsubsection{Data Collection and Analysis}
The qualitative analysis reported here focuses on the three expert participants. After each session, the instructor and the relevant access provider completed open-ended surveys reflecting on the three access conditions. After all six sessions in each study, a brief follow-up semi-structured interview was conducted with the expert participant, and the moderator recorded interview notes. The present analysis draws on the expert participants' survey responses and interview notes.

These data were analyzed inductively by the study researchers using reflexive thematic analysis \cite{braun_using_2006}. The researchers iteratively reviewed the survey responses and interview notes, generated initial codes, and developed themes across participants and modalities to identify recurring barriers, facilitators, and design implications in the in-person, remote, and AR-mediated conditions. Given the exploratory and expert-based nature of the study, the analysis focused on identifying recurring patterns across sessions and conditions rather than achieving thematic saturation or generalizable findings.

%% file: texfiles/4-findings.tex
\section{Findings}
We report findings across themes, focusing on how AR-mediated communication access reshapes accessibility labor and interactional practices among instructors, interpreters, and captioners in hands-on learning environments.

\subsection{Familiarity vs. Mediated Access}
In-person condition was consistently described as the most natural and reliable. The instructor noted that in-person interpreting \textit{“feels the most familiar,”} while interpreters described it as \textit{“least cognitively demanding.”} Captioners similarly characterized in-person captioning as \textit{“effortless and highly familiar.”} However, this condition limited visibility and introduced spatial constraints. The instructor reported, \textit{“I cannot see the student's face or work station so I don't know when they are confused or what step they are on in the experiment. That's a big disadvantage.”} The instructor also emphasized that \textit{“having another body in the room is always difficult in a science lab.”} Interpreters also experienced line-of-sight issues. Remote and AR-mediated modalities reduced these physical constraints but introduced technological mediation. Remote interpreting was described as \textit{“impersonal,”} while Zoom captioning was described as frustrating, with the captioner noting, \textit{“I definitely do not prefer to type into Zoom. It just doesn't feel that great typing into the tiny window and then having my captions disappear.”}
ARRAE balanced these trade-offs by improving access while retaining some mediation challenges.

\subsection{Visual Awareness and Task Coordination}
Across all roles, visual access to the student, workspace, and communication stream was key to effective coordination. Remote Access and ARRAE conditions improved instructor awareness. The instructor explained, 
\textit{“I really enjoy being able to see the student's face...limited view of the student's workstation...and also the interpreter. I can see when the student is confused or needs further clarification. I can also see what part of the task the student is working on. This helps me with pacing of giving directions.”}
ARRAE further enhanced this through multi-camera views: \textit{“Multiple camera views is very helpful, especially when I'm able to see what the student is doing. This helps with how I pace my instructions."} Interpreters and captioners similarly emphasized contextual awareness. The captioner noted, \textit{“I also liked how there were multiple camera angles that allowed me to see what the student was working on, which just gave me some more context which helps with captioning.”} However, visibility remained fragile. In remote setups, occlusion from materials disrupted communication, and similar issues persisted in ARRAE. These findings highlight that robust, multi-perspective visibility is essential but not fully resolved.

\subsection{Attention Integration and Interactional Flow}
Conditions differed in how they supported attention during tasks. Remote Access setups required attention shifts or movement. The instructor observed that students had to \textit{“move between the supply desk and the fume hood”} to communicate, and Zoom captions required looking away from the task. ARRAE reduced this overhead by integrating communication into the student’s field of view. The instructor noted, \textit{“the student was able to continue working while still watching the interpreter through the glasses instead of needing to shift [their] gaze to the iPad [as in the case with Zoom].”} Similarly, \textit{“[They were] able to work while reading the captions simultaneously and as an instructor, I thought that was efficient.”} Interpreters also reported reduced reliance on gaze alignment. However, ARRAE introduced coordination challenges. Instructors noted difficulty initiating interaction, requiring coordination with both interpreter and camera positioning. This suggests a need for better attention signaling mechanisms.

\subsection{Instructor Awareness and Control of Communication}
Instructor awareness of communication processes was critical for pacing and clarity. In remote captioning, instructors valued seeing captions: \textit{“I like being able to see the captions…so I know if I need to slow down.”} Similarly, visibility of the student and interpreter supported comprehension monitoring. In ARRAE, this awareness was reduced. The instructor noted, \textit{“I wish I could see the captions myself so that I know instructions are being conveyed correctly and if I need to slow down my pacing.”} While AR improved student access, it limited instructor oversight. In-person setups provided indirect awareness through environmental cues (e.g., \textit{“seeing/hearing the captionist type helped me with my pace”}). Overall, these findings indicate a trade-off between optimizing student experience and maintaining instructor control.

\subsection{Usability and Workflow Constraints}
Technology-mediated modalities introduced usability and reliability challenges. Remote Access setups were affected by audio capture issues, setup complexity, and platform limitations. For example, instructors described setup as cumbersome: \textit{“Making sure everyone except the instructor is muted and that the captioner is granted manual caption ability is just extra stuff to have to take care of before ‘class’ starts.”} ARRAE improved audio quality, contextual awareness, and interface ergonomics. The captioner noted, 
\textit{“Setup and delivery was pretty seamless. The audio was much more clear than Zoom as well.”}
However, ARRAE introduced new challenges, including system instability (\textit{“freezing of the video feed”}), limited interface flexibility, and hardware constraints. These findings highlight the need for reliable, low-friction setup and ergonomic design in AR systems.

\subsection{Persistence of Information for Task Continuity}
Persistence of communication content was critical in multi-step lab tasks. In-person captioning supported this through stable text displays. As the instructor noted, \textit{“Everything went smoothly, I noticed that the student was able to refer back to the iPad screen to verify [they] had gotten all of [their] supplies. On Zoom and with the smart glasses, there wasn't enough text that remained on the screen for [them] to do so.”} In contrast, Zoom captions were transient: \textit{“The captions…disappear quickly.”} ARRAE improved persistence by keeping captions visible within the student’s field of view. The instructor observed, \textit{“I noticed that the student was able to refer back to the captioning text still visible on the glasses while [they] gathered all of [their] required materials.”} These findings highlight the importance of persistent, integrated access to communication in supporting task continuity.

Overall, these findings reveal consistent trade-offs across modalities. In-person access offers reliability but limits visibility and scalability. Remote approaches improve awareness but introduce fragmentation and technical challenges. AR-mediated access integrates communication into ongoing activity and enhances contextual awareness, but currently limits instructor oversight and presents usability constraints. A summary of these trade-offs across modalities is provided in Table~\ref{tab:comparison}.

\begin{table}[t]
\caption{Comparison of Communication Access Modalities Across Key Themes}
\label{tab:comparison}
\centering
\small
\begin{tabular}{>{\raggedright\arraybackslash}p{2.2cm} 
                >{\raggedright\arraybackslash}p{2.9cm} 
                >{\raggedright\arraybackslash}p{2.9cm} 
                >{\raggedright\arraybackslash}p{2.9cm}}
\hline
\textbf{Themes} & \textbf{In-person} & \textbf{Remote Access} & \textbf{ARRAE} \\
\hline
Familiarity \& Reliability & High, no tech issues & Moderate, platform dependent & Moderate; emerging, some instability \\
\hline
Visual Awareness & Limited, restricted line-of-sight & Improved; single view of student/workspace & High; multi-camera views, richer context \\
\hline
Attention \& Interaction & Direct, requires co-location & Requires gaze shifts and repositioning & Integrated into task; reduced gaze shifts \\
\hline
Instructor Awareness & Indirect cues only & High; can see captions/interpreter & Limited; cannot see captions directly \\
\hline
Usability \& Setup & Simple, no setup & Moderate complexity; audio, roles, setup & Higher complexity; multi-camera positioning \\
\hline
Information Persistence & High; stable display & Low; transient captions & Moderate--high; persistent in-view captions \\
\hline
Space Constraints & High (extra body in lab) & Low & Low \\
\hline
\end{tabular}
\end{table}

%% file: texfiles/5-discussion.tex
\section{Discussion}
Our findings suggest that AR reconfigures how communication, attention, and awareness are distributed across instructors, interpreters, captioners, and students, rather than simply improving access. We interpret these findings below.

\subsection{Communication Access as Distributed Awareness}

A key contribution of this work is the reframing of communication access as \textit{distributed awareness}. Within the context of this study, effective instructional coordination depended on how well the instructor, interpreter or captioner, maintained awareness of student activity, communication flow, and task progress. In-person condition supported ecological awareness through co-presence, enabling interpreters and captioners to access environmental cues with minimal effort. However, this came at the expense of instructor awareness as it limited the instructor’s ability to monitor comprehension and progress.

Remote and AR-mediated modalities redistributed awareness through video streams. ARRAE, in particular, expanded awareness by providing simultaneous views of the student, workspace, and communication channels, supporting more informed instructional pacing. However, this redistribution was uneven. While AR improved student-facing access and provider awareness, it reduced instructor's visibility into communication outputs, particularly captions. This asymmetry highlights that increasing visibility for one role may reduce it for another, suggesting that communication access systems must be understood as shared, multi-user awareness environments rather than individual assistive tools.

\subsection{Embodied Interaction and Integration of Communication}

The findings also point to a shift in how communication is integrated into embodied activity. In traditional configurations, communication and task execution are spatially separated, requiring students to alternate attention between the task and the communication channel. This introduces interactional overhead, particularly in dynamic environments such as laboratory settings. AR-mediated access reduces this separation by embedding communication within the student’s perceptual field.  From a distributed cognition perspective, this integration reduces the need for explicit coordination but introduces new interactional challenges. Interpreters reported reduced dependence on gaze alignment, while instructors experienced difficulty initiating interaction without shared physical cues. These findings suggest that while AR supports continuous access, it alters the mechanisms through which participants establish attention and coordination. 

\subsection{Trade-offs Between Access, Control, and Coordination}

The findings reveal persistent trade-offs between improving access, maintaining instructional control, and supporting coordination. First, enhancing student access through AR can reduce instructor control over communication processes. In contrast to remote captioning, where instructors could monitor captions directly, ARRAE limited this visibility. This suggests a tension between student-centered access and instructor oversight of communication accuracy and pacing. Second, while AR enhances contextual awareness through multi-perspective views, it introduces coordination overhead. Participants reported challenges related to camera positioning, system setup, and managing occlusions. This reflects a broader pattern where technological augmentation redistributes coordination work rather than eliminating it. Third, the persistence of information played a key role in supporting task continuity. In-person captioning enabled students to revisit information easily, while Zoom captions were described as transient. ARRAE addressed this by allowing captions to remain visible within the student’s workspace, supporting continuous engagement without repeated clarification. However, this was limited by the space available in the captioning container seen within the AR view. 

\subsection{Implications for Experiential Learning Contexts}

Collectively, the findings from this study offer initial insights into technology-mediated interaction in experiential learning environments. In laboratory settings, where perception, action, and communication are tightly coupled, our findings suggest that AR-mediated systems may support this coupling by integrating communication into ongoing activity and enhancing contextual awareness. At the same time, the effectiveness of such systems appears to depend on how well they align with the social and spatial organization of the classroom. Within the context of this study, AR-mediated access did not function as a replacement for existing modalities, but rather as part of a hybrid configuration in which different approaches afforded different forms of awareness, control, and interaction.

%% file: texfiles/6-limitations-futurework.tex
\section{Limitations and Future Work}
This work should be interpreted as an exploratory study of AR-mediated communication access in experiential higher education. The study involved a single instructor, interpreter, and captioner, enabling a close examination of expert workflows and modality-specific trade-offs, but not supporting statistical generalization. The findings are therefore intended to inform early-stage design and theory-building for communication access in hands-on instructional settings. Replication with multiple professionals in longitudinal studies is needed to assess whether the observed patterns hold across different instructors and access service providers over time, and how increasing familiarity may reshape coordination demands, usability, and instructional fit. Additionally, some challenges observed in the ARRAE condition may reflect the current maturity of the prototype, as well as the broader demands of AR-mediated communication access.

Future work should also move beyond technical feasibility toward professional efficacy in experiential learning environments. Key priorities include addressing the fractured interactional feedback reported in AR-mediated conditions, particularly reduced access to non-verbal and backchanneling cues used to gauge comprehension, pacing, and communicative alignment—and refining the system in line with the requirements identified in this study. These include role-specific visual views, shared awareness features for instructors, improved persistence and review of captioning content, more flexible interface controls, and more dependable audio and video performance. Future studies should further examine these issues across a broader set of instructors, students, interpreters, and captioners, and across diverse experiential tasks, to assess how these patterns generalize under varied user demands and whether AR-mediated access can become dependable in routine practice.

Finally, as a broader future direction, we plan to explore the presentation of additional elements of speech, such as affective captions, which have been shown to enhance user experience~\cite{fuzzyCHI, caluaCHI23, captionRoyaleCHI, hapticCaptioningCHI23, de2025tactile, cucapASSETS2025}. However, it is critical to examine how these features can be effectively delivered from a service-provider perspective, as well as their impact on users’ experiences. Accordingly, our future work will investigate both the design and implications of such approaches.

%% file: texfiles/7-conclusion.tex
\section{Conclusion} 
Accessibility in experiential higher education is not simply a matter of delivering communication access; it also depends on how that access supports relationships among students, instructors, and access providers during active task work. This exploratory study suggests that AR-mediated support can help restore forms of spatial and contextual awareness that are often diminished in conventional remote access. In particular, ARRAE showed promise in improving shared visual awareness, reducing gaze shifting, and better integrating interpreting and captioning into hands-on laboratory activity. At the same time, in-person support remained the most familiar and dependable modality, while the AR condition introduced new demands related to positioning, interface flexibility, and system reliability. Overall, these findings suggest that AR-mediated access in experiential higher education may be most effective when designed not simply as a wearable replacement for traditional interfaces, but as a shared system that supports the distinct yet interconnected needs of students, instructors, interpreters, and captioners through shared awareness, feedback loops, and reliable coordination.

%% file: texfiles/8-credits.tex
\begin{credits}
\subsubsection{\ackname} The contents of this paper were partially developed under a grant from the National Institute on Disability, Independent Living, and Rehabilitation Research (NIDILRR grant number 90IFRE0083). NIDILRR is a Center within the Administration for Community Living (ACL), Department of Health and Human Services (HHS). The contents of this paper do not necessarily represent the policy of NIDILRR, ACL, or HHS, and you should not assume endorsement by the Federal Government. The authors thank Ms. Wendy Dannels for her guidance and valuable contributions to the development of this study. 
\end{credits}

%% file: bib/main.bib
@online{typewell_transcription,
	title = {{TypeWell} {\textbar} Real-Time Transcription Services and Professional Training for Accessibility},
	url = {https://typewell.com/},
	titleaddon = {{TypeWell}},
	urldate = {2026-01-26},
	langid = {american},
}

@incollection{mayer_spatial_contiguity_principle_2009,
	location = {Cambridge},
	edition = {2},
	title = {Spatial Contiguity Principle},
	isbn = {978-1-107-79037-7},
	url = {https://www.cambridge.org/core/books/multimedia-learning/spatial-contiguity-principle/B9B79EDC777C375C7ED410B82EF80247},
	doi = {10.1017/CBO9780511811678.010},
	pages = {135--152},
	booktitle = {Multimedia Learning},
	publisher = {Cambridge University Press},
	editor = {Mayer, Richard E.},
	urldate = {2026-01-26},
	date = {2009},
}

@article{braun_using_2006,
	title = {Using thematic analysis in psychology},
	volume = {3},
	issn = {1478-0887},
	url = {https://doi.org/10.1191/1478088706qp063oa},
	doi = {10.1191/1478088706qp063oa},
	pages = {77--101},
	number = {2},
	journaltitle = {Qualitative Research in Psychology},
	publisher = {Routledge},
	author = {Braun, Virginia and Clarke, Victoria},
	urldate = {2026-01-26},
	date = {2006-01-01},
	note = {\_eprint: https://doi.org/10.1191/1478088706qp063oa},
}

@inproceedings{fuzzyCHI,
author = {de Lacerda Pataca, Calu\~{a} and Patterson, Stephanie and Peiris, Roshan and Huenerfauth, Matt},
title = {Fuzzy Feelings: Arousal’s Interpretive Noise and the Case for Acoustic-Based Haptics},
year = {2026},
isbn = {9798400722783},
publisher = {Association for Computing Machinery},
booktitle = {Proceedings of the CHI Conference on Human Factors in Computing Systems},
address = {New York, NY, USA},
url = {https://doi.org/10.1145/3772318.3793421},
doi = {10.1145/3772318.3793421},
numpages = {13},
location = {Barcelona, Spain},
series = {CHI '26}
}

@inproceedings{cucapASSETS2025,
author = {de Lacerda Pataca, Calua and Ahn, SooYeon and Yoo, Suhyeon and Kim, JooYeong and Truong, Khai N. and Hong, Jin-Hyuk and Peiris, Roshan and Huenerfauth, Matt},
title = {CuCap: Comparative Analysis of Customized Captioning between North American and South Korean d/Deaf and Hard-of-Hearing Users},
year = {2025},
publisher = {Association for Computing Machinery},
address = {New York, NY, USA},
booktitle = {The 27th International ACM SIGACCESS Conference on Computers and Accessibility - Conditionally Accepted},
location = {Denver, Colorado},
series = {ASSETS '25}
}

@inproceedings{de2025tactile,
author = {de Lacerda Pataca, Calu\~{a} and Hassan, Saad and May, Lloyd and Olson, Michelle M and D'aurio, Toni and Peiris, Roshan L and Huenerfauth, Matt},
title = {Tactile Emotions: Multimodal Affective Captioning with Haptics Improves Narrative Engagement for d/Deaf and Hard-of-Hearing Viewers},
year = {2025},
isbn = {9798400713941},
publisher = {Association for Computing Machinery},
address = {New York, NY, USA},
url = {https://doi.org/10.1145/3706598.3713304},
doi = {10.1145/3706598.3713304},
booktitle = {Proceedings of the 2025 CHI Conference on Human Factors in Computing Systems},
articleno = {68},
numpages = {17},
location = {
},
series = {CHI '25}
}

@inproceedings{captionRoyaleCHI,
author = {de Lacerda Pataca, Calu\~{a} and Hassan, Saad and Tinker, Nathan and Peiris, Roshan Lalintha and Huenerfauth, Matt},
title = {Caption Royale: Exploring the Design Space of Affective Captions from the Perspective of Deaf and Hard-of-Hearing Individuals},
year = {2024},
isbn = {9798400703300},
publisher = {Association for Computing Machinery},
address = {New York, NY, USA},
url = {https://doi.org/10.1145/3613904.3642258},
doi = {10.1145/3613904.3642258},
booktitle = {Proceedings of the CHI Conference on Human Factors in Computing Systems},
articleno = {899},
numpages = {17},
series = {CHI '24}
}

@inproceedings{hapticCaptioningCHI23,
author = {Wang, Yiwen and Li, Ziming and Chelladurai, Pratheep Kumar and Dannels, Wendy and Oh, Tae and Peiris, Roshan L},
title = {Haptic-Captioning: Using Audio-Haptic Interfaces to Enhance Speaker Indication in Real-Time Captions for Deaf and Hard-of-Hearing Viewers},
year = {2023},
isbn = {9781450394215},
publisher = {Association for Computing Machinery},
address = {New York, NY, USA},
url = {https://doi.org/10.1145/3544548.3581076},
doi = {10.1145/3544548.3581076},
booktitle = {Proceedings of the 2023 CHI Conference on Human Factors in Computing Systems},
articleno = {781},
numpages = {14},
location = {Hamburg, Germany},
series = {CHI '23}
}

@inproceedings{caluaCHI23,
author = {de Lacerda Pataca, Calu\~{a} and Watkins, Matthew and Peiris, Roshan and Lee, Sooyeon and Huenerfauth, Matt},
title = {Visualization of Speech Prosody and Emotion in Captions: Accessibility for Deaf and Hard-of-Hearing Users},
year = {2023},
isbn = {9781450394215},
publisher = {Association for Computing Machinery},
address = {New York, NY, USA},
url = {https://doi.org/10.1145/3544548.3581511},
doi = {10.1145/3544548.3581511},
booktitle = {Proceedings of the 2023 CHI Conference on Human Factors in Computing Systems},
articleno = {831},
numpages = {15},
location = {Hamburg, Germany},
series = {CHI '23}
}

@online{rid_resources_2022,
	title = {Resources - Registry of Interpreters for the Deaf, Inc.},
	url = {https://rid.org/resources/},
	urldate = {2026-01-26},
	date = {2022-08-01},
	langid = {american},
}

@incollection{hoza_team_2022,
	title = {Team interpreting},
	booktitle = {The Routledge Handbook of Sign Language Translation and Interpreting},
	publisher = {Routledge},
	author = {Hoza, Jack},
	date = {2022},
	note = {Num Pages: 17},
}

@inproceedings{miller_use_2017,
	location = {New York, {NY}, {USA}},
	title = {The Use of Smart Glasses for Lecture Comprehension by Deaf and Hard of Hearing Students},
	isbn = {978-1-4503-4656-6},
	url = {https://dl.acm.org/doi/10.1145/3027063.3053117},
	doi = {10.1145/3027063.3053117},
	series = {{CHI} {EA} '17},
	pages = {1909--1915},
	booktitle = {Proceedings of the 2017 {CHI} Conference Extended Abstracts on Human Factors in Computing Systems},
	publisher = {Association for Computing Machinery},
	author = {Miller, Ashley and Malasig, Joan and Castro, Brenda and Hanson, Vicki L. and Nicolau, Hugo and Brandão, Alessandra},
	urldate = {2026-01-26},
	date = {2017-05-06},
}

@online{vuzix_blade2,
	title = {Vuzix Blade 2™ Smart Glasses},
	url = {https://www.vuzix.com/products/vuzix-blade-2-smart-glasses},
	titleaddon = {Vuzix Corporation},
	urldate = {2026-01-26},
	langid = {english},
}

@article{smithchemical2016,
	title = {Chemical and biological research with deaf and hard-of-hearing students and professionals: Ensuring a safe and successful laboratory environment},
	volume = {23},
	issn = {1871-5532},
	url = {https://doi.org/10.1016/j.jchas.2015.03.002},
	doi = {10.1016/j.jchas.2015.03.002},
	shorttitle = {Chemical and biological research with deaf and hard-of-hearing students and professionals},
	pages = {24--31},
	number = {1},
	journaltitle = {Journal of Chemical Health \& Safety},
	shortjournal = {J. Chem. Health Saf.},
	publisher = {American Chemical Society},
	author = {Smith, Susan B. and Ross, Annemarie D. and Pagano, Todd},
	urldate = {2026-01-26},
	date = {2016-01-01},
}

@inproceedings{mathew_access_2022,
	location = {New York, {NY}, {USA}},
	title = {Access on Demand: Real-time, Multi-modal Accessibility for the Deaf and Hard-of-Hearing based on Augmented Reality},
	isbn = {978-1-4503-9258-7},
	url = {https://dl.acm.org/doi/10.1145/3517428.3551352},
	doi = {10.1145/3517428.3551352},
	series = {{ASSETS} '22},
	shorttitle = {Access on Demand},
	pages = {1--6},
	booktitle = {Proceedings of the 24th International {ACM} {SIGACCESS} Conference on Computers and Accessibility},
	publisher = {Association for Computing Machinery},
	author = {Mathew, Roshan and Mak, Brian and Dannels, Wendy},
	urldate = {2026-01-26},
	date = {2022-10-22},
}

@misc{obsbot_tail_air,
	title = {{OBSBOT} {Tail} {Air}: {AI}-{Powered} {4K} {PTZ} {Streaming} {Camera}},
	url = {https://www.obsbot.com/obsbot-tail-air-streaming-camera},
	urldate = {2025-09-11},
}

@misc{tenveo_vhd10u,
	title = {{TEVO}-{VHD10U} {USB} {HDMI} {10X} optical {Zoom} {Conference} {Camera}},
	url = {https://www.tenveo.com/product/ptz-camera/418.html},
	urldate = {2025-09-12},
}

@misc{obsbot_tiny,
	title = {{OBSBOT} {Tiny} {4K} - {AI}-{Powered} {AI} {Tracking} {PTZ} {4K} {Webcam}},
	url = {https://www.obsbot.com/obsbot-tiny-4k-webcam},
	language = {en-US},
	urldate = {2025-09-12},
	journal = {OBSBOT - The Best Auto-Tracking Webcam},
}

@inproceedings{mathew_improving_2025,
	location = {New York, {NY}, {USA}},
	title = {Improving the User Experience for Sign Language Interpreters in Remote Interpretation for Small Group Settings},
	isbn = {979-8-4007-1882-3},
	url = {https://dl.acm.org/doi/10.1145/3744257.3744261},
	doi = {10.1145/3744257.3744261},
	series = {W4A '25},
	pages = {51--62},
	booktitle = {Proceedings of the 22nd International Web for All Conference},
	publisher = {Association for Computing Machinery},
	author = {Mathew, Roshan and Dannels, Wendy},
	urldate = {2026-01-11},
	date = {2025-10-15},
}

@inproceedings{mathew_experiential_2026,
	location = {New York, {NY}, {USA}},
	title = {Evaluating the Feasibility of Augmented Reality to Support Communication Access for Deaf Students in Experiential Higher Education Contexts.},
	isbn = {979-8-4007-2372-8},
	url = {https://doi.org/10.1145/3800424.3800450},
	doi = {10.1145/3800424.3800450},
	series = {W4A '26},
	booktitle = {Proceedings of the 23rd International Web for All Conference},
	publisher = {Association for Computing Machinery},
	author = {Mathew, Roshan and Peiris, Roshan L.},
}
